# SOE's ESG Performance on Financial Flexibility: The Evidence from the Hong Kong Stock Market

LI Yan

1155191643@link.cuhk.edu.hk

The Chinese University of Hong Kong

**Abstract:**
As the global economic environment becomes increasingly unstable, enhancing financial flexibility to cope with risks has become the consensus of many companies. At the same time, environmental, social, and governance (ESG) performance may be one of the effective ways. We studied the impact of a firm's ESG performance on its financial flexibility with a sample of companies listed on the Hong Kong stock market from 2018 to 2022. The empirical results show that good environmental, social and governance performance can significantly improve a firm's financial flexibility. In addition, this paper also finds that the influence of ESG performance on financial flexibility is weak for state-owned enterprises due to the influence of governance structure and market characteristics. Finally, the further analysis shows that there is a mediating role played by financing constraints in this process. This study can provide background information for state-owned enterprises' governance, information disclosure, and corporate operations. It also has guiding significance for relevant investors, management and officials.

1. Introduction

In recent years, the world economy has continued to be volatile, accompanied by various unfavorable conditions, and the risks of doing business have subsequently increased significantly. For companies, enhancing sustainable development and resilience to risks becomes even more critical in times of crisis.

Financial flexibility is the ability of a firm to obtain and restructure financing at a low cost, and firms can often enhance financial flexibility by deleveraging and increasing cash holdings. (Gamba & Triantis, 2008, DeAngelo et al., 2017). Financial flexibility not only mitigates the problem of underinvestment by firms to facilitate their growth. More importantly, it can help avoid the firm's financial difficulties and enhance its risk resistance. Firms with greater flexibility also tend to perform better in crises (Arslan-Ayaydin et al., 2013). In conclusion, financial flexibility facilitates significant mitigation of business risks and is very important in today's economic environment.

ESG is a system including environmental, social, and governance indicators. It is a non-traditional performance concept and corporate evaluation standard, which is widely used to evaluate the sustainability level of firms. In recent years, along with the growing interest in sustainable investment, the demand for ESG information has also increased rapidly (Cohen et al., 2015). Good ESG performance can send positive signals to the external world, representing a stronger level of sustainable development for the firm. Therefore, it is more likely to be favored by stakeholders such as investors and consumers, which can improve the firm's financial situation.

There are plenty of previous studies on the impact of ESG on firms, but most of them focus on the impact of ESG on firms' performance and value, with few studies focusing on the effects of ESG performance on financial flexibility (Tsang et al., 2023). In addition, past studies have found that the effect of ESG performance is linked to market infrastructure and firm characteristics (Li et al., 2021), while the extant samples of studies on the impact of ESG on financial flexibility are almost from a single economy. Finally, extant studies on the impact of ESG on financial flexibility focus more on its mechanism of action and do not pay attention to examining firms with different characteristics. In conclusion, the effect of ESG performance on firms' financial flexibility is unclear and lacks details.

Hong Kong is a place where East and West intermingle economically and culturally. On the one hand, Hong Kong has long served as a communication window between China and the world. A large number of Chinese companies are listed in Hong Kong stocks, while the Chinese capital market and the ESG construction of Chinese companies are usually not well-developed. On the other hand, as one of the world's financial centers, Hong Kong's ESG infrastructure is relatively more developed. The characteristics of Hong Kong provide a good sample for this study: analyzing Hong Kong companies can make up for the limitation that the samples of previous studies come from a single economy. It also allows for separate testing of firms with different characteristics to gain a more detailed understanding.

To address the above issues and conditions, this paper analyzes and examines the relationship between ESG performance and firms' financial flexibility based on the stakeholder theory and signaling theory, with samples from the Hong Kong stock market from 2018 to 2022. It is found that good ESG performance can send positive signals and alleviate the information asymmetry problem, which is also in line with the interest correlator theory, thus making it easier to obtain financing support. In addition, we conducted a separate examination for the SOEs in our sample and found that the effect of ESG on financial flexibility is weaker on SOEs.

First, as most of the extant studies focus on the impact of ESG on firm value, our study supports the positive effect of ESG performance on financial flexibility, further enriching the relevant literature. Second, among similar studies, this paper conducts a separate examination of the category of SOEs for the first time, further investigating the differences in the impact of ESG. Finally, we also analyze the path of ESG's impact on financial flexibility from the perspective of financing constraints.

The paper is organized as follows: the first section is an introduction, followed by a literature review and two hypotheses. The third section describes the sample and the research model. The fourth section presents and discusses the results. The fifth section addresses the impact pathways for further analysis. The sixth section points out the conclusions, practical significance, and recognizes the limitations.

## 2. Literature Review and Hypotheses Development
### 2.1. ESG Overview

As more and more attention is paid to sustainable development and social responsibility, how to quantify and assess them becomes crucial. ESG, which consists of environmental, social and governance components, first appeared in the United Nations' Responsible Investment Choices report in 2004, and is now widely used in assessing a company's overall social responsibility performance. As ESG is a non-financial indicator, it does not have a standardized measure, but in general, the environmental component measures a company's efforts to go green and reduce carbon emissions, and the social component measures a company's performance in areas such as human rights. Finally, the governance component shows the level of corporate management structure and management efficiency.

Although ESG was proposed late, research on the relationship between ESG and firm performance has been abundant. Past studies have focused on the impact of ESG on financial performance (Friede et al., 2015), financial decision-making (Eccles et al., 2018), firm market capitalization (Zhou et al., 2022), and return on investment (Jain et al., 2019). In addition, the studies of ESG impacts are quite detailed and extensive, such as different industries (Ionescu et al., 2019; Ersoy et al., 2022), different characteristics (Yoon et al., 2018; Ciciretti et al., 2023), and so on.

The impact of ESG still needs a unified and definitive picture. A considerable number of scholars believe that the most crucial goal of a firm is to maximize profit (Friedman, 2007). According to the theory, there is hardly any economic benefit from the inputs of ESG performance; instead, a lot of additional costs are incurred, which are harmful to the operation of the firm. In contrast, according to the stakeholder theory, ESG benefits various stakeholders, such as customers, society, and government, and can contribute to business development (Kumar, 2023; Freeman et al., 2021). At the same time, it has been argued that ESG performance does not make a material difference to firms (Khan et al., 2016).

*2.2. ESG Performance and Financial Flexibility*

Financial flexibility represents a firm's ability to face unexpected shocks to cash flows and investments (Bancel & Mittoo, 2004). Financial flexibility is mainly from low leverage and higher cash holdings, which can help maintain the flexibility of the firm's operations. In the event of a crisis, firms with high financial flexibility tend to be more moderately affected (Bancel & Mittoo, 2011).

Adverse selection and moral hazard are two information asymmetry problems in financial markets, which increase the difficulty of financing. Signaling theory helps to describe the behavior of both parties when they get different information (Connelly et al.). According to signaling theory, on the one hand, ESG disclosure improves external understanding of the firm and reduces information asymmetry (Ching & Gerab, 2017). On the other hand, good ESG performance releases good signals to the outside world and builds a better image, and in this way enhances investor confidence and willingness to invest, thereby increasing a firm's cash flow and financial flexibility. For example, ESG investing is increasingly recognized by investors as it implies higher portfolio performance and lower risk (Broadstock et al., 2021).

From a consumer perspective, good ESG performance can enhance a product's brand value and image, which improves competitiveness and consumer favor (Lee et al., 2022). Additionally, with consumers increasingly concerned about green consumption (Griskevicius et al., 2010), ESG and its sustainability concept will make products popular and easier to sell. In short, ESG performance can improve firms' profitability by boosting product sales, thereby increasing profit, cash flow and enhancing financial flexibility.

Then, according to stakeholder theory, sustainable development can only be achieved by a firm that is accountable to all its stakeholders. Stakeholders may include internal stakeholders, such as shareholders and employees, as well as external stakeholders, including society and government (Parmar et al., 2010). The stakeholder theory opposes the shareholder supremacy theory and emphasizes that firms should balance social values and profitability. Thus, it appears that companies with good ESG performance represent a concern for ethical values and are more likely to achieve sustainable development. As a result, they are also more supportive in the face of an uncertain environment, with more excellent risk response capabilities and financial flexibility.

In summary, we argue that good ESG performance can enhance firms' financial flexibility and propose Hypothesis 1 (H1):

**Hypothesis 1 (H1):** *Good ESG performance contributes to the financial flexibility of an enterprise.*

### 2.3. Weaker Effect for SOEs

China has the world's largest system of state-owned assets, and the dominance of the public sector economy notably characterizes its economic system. There are a large number of state-owned enterprises (SOEs) in China's capital market, which are the economic foundation of the Chinese government and the Communist Party (Zahid et al., 2023). Since the 1980s, China has carried out several reforms, and currently, SOEs have formed a dual governance system where corporate governance and political governance coexist (Wang, 2014). The corporate governance system focuses on maximizing shareholders' interests, while the political governance system takes on governmental and political tasks. In addition, government officials who manage SOEs also tend to pay more attention to political tasks and are less sensitive to ESG signals. In short, the incentives for SOEs to participate in ESG are not yet well developed (He et al., 2023).

Compared to general firms, on the one hand, SOEs have the full support of the government, especially in terms of financial resources. On the other hand, Chinese banks (most of which are state-owned) tend to provide easier credit support to SOEs due to the government's credit endorsement, and thus, Chinese SOEs often have easier access to external financing (Song et al., 2011). The above reasons lead to SOEs' stronger financing ability and higher financial flexibility on their own, and they are less dependent on the positive effect of ESG performance on financial flexibility. To summarize, we believe that the impact of ESG on financial Flexibility is weaker for state-owned enterprises than for firms in general and propose Hypothesis 2 (H2):

**Hypothesis 2 (H2):** *The impact of ESG on financial Flexibility is weaker for state-owned enterprises than for firms in general.*

## 3. Research Methods
### 3.1. Sample Selection and Data Source

This study selected firms listed on the Hong Kong stock market from 2018 to 2022 to investigate the impact of ESG performance on firm financial flexibility. The ESG score, financial flexibility, and other variables were sourced from the Refinitiv database. To ensure the validity of the research data, the following methods were employed for data screening and processing: (1) All the samples are constituent stocks of the Hang Seng Composite Index to ensure the representativeness of the samples. (2) Firms with missing or abnormal variable data were excluded. (3) The background types of firms were identified. The final sample consisted of 969 panel data points from 210 firms.

### 3.2. Variable Setting

Financial flexibility (FF): Financial flexibility is the dependent variable in this study, which is used to measure a firm's financial capacity to cope with risks in an uncertain environment (Rapp et al., 2014). There is still no fixed standard for measuring financial flexibility. Based on previous research, cash is an essential source of financial flexibility (Esin, 2015). Therefore, this study selects operating cash flow divided by total assets to measure financial flexibility, highlighting the firm's ability to manage operational risks effectively.

ESG performance (ESG): ESG performance is the independent variable in this study, and it is measured using ESG scores obtained from Refinitiv. Refinitiv, with over 40,000 employees in 190 countries, operates the world's largest ESG content collection business. They regularly maintain and calculate ESG scores for over 15,000 companies globally. Refinitiv's ESG scores measure companies' ESG performance based on publicly available and verifiable reporting data. It collects and calculates over 630 enterprise-level ESG indicators, with the most comparable and significant 186 indicators supporting the overall assessment and scoring process for companies within each industry. The formulation of these indicators considers the varying comparability, impact, data availability, and industry relevance across different industry groups. These indicators are categorized into ten categories, forming three pillar scores and the final ESG score, which reflect the company's ESG performance, commitments, and effectiveness based on publicly reported information.

Drawing on previous research on financial flexibility (Zhang & Liu, 2022; Naseer et al., 2023), In order to enhance the validity of the results, this study selects four control variables, including return on net assets, firm size, book-to-market ratio, and EBITDA-to-interest expenses ratio.

In addition, in order to test H2, we define firms whose real controllers are the Chinese government (excluding Hong Kong, Macao, and Taiwan) as state-owned enterprises.

The explanation and set of variables are in Table 1.

**Table 1. Variables Description**

| Symbol | Variable | Variable Definition |
|---|---|---|
| ESG | ESG performance | According to ESG Score from Refinitiv |
| FF | Financial Flexibility | Operating Net Cash Flow / Total Asset |
| FC | Financing Constraint | KZ index |
| Size | Firm Size | The logarithm of total asset |
| ROE | Return on Equity | Net Income / Shareholder Equity |
| EBIN | EBITDA interest coverage | EBITDA / Interest Expenditure |
| BM | Book-to-Market Value | Book Value / Market Value |
| SOE | Property rights | If the firm is a state-owned enterprise, SOE=1 |
| IND | Industry-fixed effect | Industry dummies |
| YEAR | Year-fixed effect | Year dummies |

### 3.3. Methodological Remarks

In order to verify the relationship between the firm's ESG performance and the firm's financial flexibility, this paper adopts a two-way fixed-effects model of controlling the industry and year to establish Equation (1):

$$FF_{i,t} = \beta_0 + \beta_1 ESG_{i,t} + \beta_2 ROE_{i,t} + \beta_3 Size_{i,t}$$
$$+\beta_4 EBIN_{i,t} + \beta_5 BM_{i,t} + \sum IND + \sum YEAR + \varepsilon_{i,t} \quad (1)$$

In this equation, $FF$ is the financial flexibility of the explained variable, $ESG$ is the ESG performance of the explanatory variable, and there are also some control variables, including Return on Equity ($ROE$), EBITDA interest coverage ($EBIN$), Book-to-Market Value ($BM$) and Firm Size ($Size$). Detailed definitions of these variables are given in Table 1.

Then, we further investigate whether the impact of ESG performance on financial flexibility differs across firms. Firstly, we hypothesize that the impact of ESG on financial flexibility is less significant for state-owned enterprises compared to other listed companies in the Hong Kong stock market. For this proposal, we test this hypothesis by examining the coefficients of the interaction terms of ESG performance and state-owned enterprise dummy variables. Specifically, based on Equation (1), we add the interaction term between ESG scores and SOE dummy variables and establish Equation (2):

$$FF_{i,t} = \beta_0 + \beta_1 ESG_{i,t} + \beta_2 ROE_{i,t} + \beta_3 Size_{i,t} + \beta_4 EBIN_{i,t}$$
$$+\beta_5 BM_{i,t} + \beta_6 ESG_{i,t} * SOE + \sum IND + \sum YEAR + \varepsilon_{i,t} \quad (2)$$

In equation (2), if the firm is a State-owned enterprise (SOE), the indicator variable $SOE$ is equal to 1; otherwise, it is equal to 0. The coefficient $\beta_1$ represents the effect of ESG performance on the financial flexibility of normal listed firms. The coefficient $\beta_6$ of the interaction variable represents the additional effect of ESG performance on the financial flexibility of SOEs.

To summarize. Equation (1) is used to test the positive effect of ESG performance on the financial flexibility of Hong Kong-listed firms. Equation (2) is used to test the weaker impact of ESG performance on SOEs. Equations (1) to (2) will be analyzed by examining the significance of the coefficients to verify whether they hold.

**4. Results and Discussions**
*4.1. Descriptive Statistics*

Descriptive statistics are shown in Table 2, which summarizes the distribution and characteristics of each variable analyzed in the study. As shown in the results, the Mean and Median of Financial Flexibility (FF) are 0.058 and 0.048, and the Mean and Median of ESG Performance (ESG) are 54.047 and 54.639. The Mean and Median are relatively close to each other, representing that FF and ESG approximately conform to the normal distribution. The maximum and minimum values for FF are 0.571 and -0.373. For ESG, the maximum and minimum values are 89.028 and 14.485, representing a significant difference in the sample distribution and a high degree of differentiation.

**Table 2. Descriptive Statistics**

| Var Name | Obs | Mean | SD | Min | Median | Max |
| --- | --- | --- | --- | --- | --- | --- |
| FF | 969 | 0.058 | 0.075 | -0.373 | 0.048 | 0.571 |
| ESG | 969 | 54.047 | 12.461 | 14.485 | 54.639 | 89.028 |
| ROE | 969 | 5.589 | 80.783 | -2170.186 | 8.902 | 288.522 |
| Size | 969 | 9.802 | 1.774 | 3.872 | 9.706 | 16.220 |
| EBIN | 969 | 34.707 | 292.449 | -7522.017 | 8.971 | 2975.000 |
| BM | 969 | 1.437 | 3.426 | 0.001 | 0.551 | 78.900 |
| FC | 969 | -0.953 | 1.718 | -15.573 | 1.139 | 22.382 |

*4.2. Regression Result Analysis*

4.2.1. The Impact of ESG Performance on Financial Flexibility

Columns (1) to (2) of Table 3, respectively, demonstrate the results of the regressions from Equation (1) to Equation (2). In particular, the first column shows the relationship between financial flexibility and ESG performance. The second column shows the results of the additional effects of ESG on the financial flexibility of SOEs.

The coefficient of ESG in the first column is positive, about 0.001, which is significant at the 1% level, representing that good ESG performance can significantly improve firms' financial flexibility listed in the Hong Kong stock market, and Hypothesis 1 holds.

**Table 3. Regression Results**

| | (1) | (2) |
|---|---|---|
| | FF | FF |
| ESG | 0.001*** | 0.001*** |
| | (2.79) | (2.89) |
| ROE | 0.000*** | 0.000*** |
| | (3.21) | (3.21) |
| Size | 0.002 | 0.002 |
| | (0.97) | (1.16) |
| EBIN | 0.000*** | 0.000*** |
| | (2.91) | (3.01) |
| BM | 0.003*** | 0.003*** |
| | (5.11) | (5.13) |
| IND/YEAR | control | control |
| ESG∗SOE | | -0.000* |
| | | (-1.95) |
| _cons | 0.072*** | 0.072*** |
| | (3.47) | (3.49) |
| N | 941 | 941 |
| $R^2$ | 0.261 | 0.264 |

4.2.2. The effect of ESG performance on Financial Flexibility: SOEs

Column (2) of Table 3 shows the effect of the ESG performance of SOEs on their financial flexibility, from which it can be seen that the regression coefficient of the interaction term is negative and holds a 10% significance. This indicates that the effect of ESG performance of SOEs on their financial flexibility is weaker, and Hypothesis 2 holds.

*4.3. Endogeneity Mitigation: PSM Regression*

While good ESG performance promotes firms' financial flexibility, greater financial flexibility permits firms to make larger ESG investments and disclosures, thus promoting ESG performance. Such mutual causation represents an endogeneity problem for the conclusion, so we use PSM regression to mitigate the endogeneity problem.

The principle of propensity score matching is to use the propensity score values to find individuals in the treatment group from the control group with the same or similar background characteristics to serve as controls, thus minimizing the interference of other confounding factors and mitigating the endogeneity problem (Caliendo & Kopeinig, 2008).

In this paper, we divided the treatment and control groups based on the mean value of the ESG index, assigning 1 if ESG is greater than the mean value and 0 otherwise. The above control variables were selected as matching variables, and the individuals in the control and treatment groups were matched by "1:1 caliper matching (Caliper range is 0.1)" and logit regression.

The matching results of Equations (1) and (2) are shown in Table 4, where the biases are less than 10% for all variables after matching. The corresponding average treatment effects (ATT) are 2.16 and 2.14, respectively, which are significant at a 5% level. These data indicate that there is no systematic difference between the control and treatment groups and pass the balance test.

**Table 4. Balance test for PSM of Equations (1) and (2)**

| Variable | Unmatched or Matched | (1) %bias | T test | (2) %bias | T test |
|---|---|---|---|---|---|
| ROE | U | -6.5 | -0.99 | -6.6 | -1.00 |
|  | M | 4.9 | 2.33 | 5.0 | 2.42 |
| Size | U | 10.6 | 1.63 | 12.1 | 1.85 |
|  | M | -7.0 | -0.93 | -2.9 | -0.38 |
| EBIN | U | -7.9 | -1.21 | -8.1 | -1.23 |
|  | M | -0.7 | -0.16 | -2.7 | -0.60 |
| BM | U | 6.2 | 0.95 | 5.8 | 0.88 |
|  | M | -2.8 | -0.7 | -1.0 | -0.25 |
| IND/YEAR |  | control | control | control | control |
| ESG*SOE | U |  |  | 14.0 | 2.13 |
|  | M |  |  | -7.3 | -1.02 |

We rerun the regression using the matching results, and the results are shown in Table 5. The significance of the coefficients of ESG in column (1) decreases slightly to 5%. The significance of the coefficient on the ESG and SOE interaction term in column (2) rises to 5%, and the other results are consistent with previous results. In summary, both two hypotheses pass the endogeneity test.

**Table 5. PSM Regression Results**

|  | (1) FF | (2) FF |
|---|---|---|
| ESG | 0.001** | 0.001*** |
|  | (2.25) | (2.74) |
| ROE | 0.000 | 0.000 |
|  | (1.08) | (1.14) |
| Size | 0.004 | 0.009*** |
|  | (1.47) | (2.83) |
| EBIN | 0.000* | 0.000** |
|  | (1.76) | (2.00) |
| BM | 0.002** | 0.003*** |
|  | (2.37) | (3.38) |
| IND/YEAR | control | control |
| ESG*SOE | | -0.000** |
| | | (-2.14) |
| _cons | -0.022 | -0.040 |
| | (-0.40) | (-0.72) |
| N | 229 | 221 |
| $R^2$ | 0.394 | 0.379 |

### 4.4. Robustness Test: Change Measure of ESG

First, since there is currently no standardized measure of ESG performance, the same company may have different ESG scores under different evaluation systems. Second, the Refinitiv database selected for this paper uses a percentile ESG score, but a significant portion of other current ESG evaluations use a rating system, which is much weaker in differentiation (Dorfleitner et al., 2015). Finally, given that firms with low ESG performance are less invested in ESG disclosure and less likely to disclose, our current ESG sample may be skewed above the overall level. As demonstrated in the descriptive statistics, where the mean and median are higher than 50. In summary, there is a need for robustness testing by changing the measure of ESG performance to improve the persuasiveness of the findings.

Based on the above reasons, we compute the Z-score ((Score-Mean)/Standard deviation) of the sample ESG performance based on the mean and standard deviation (Sd) of the ESG scores, thus realizing the categorization process. Specifically, those within one Sd above the sample are assigned a value of 70, and those with one to two Sd are assigned a value of 90. those within one Sd below the sample are assigned a value of 30, and those with one to two Sd are assigned a value of 10. This not only simulates a rating system for other ESG evaluation methodology, but also allows for a more balanced data distribution.

After changing the measure of ESG performance and re-running the regression, the results are shown in Table 6. The positivity and negativity of all the coefficients remain unchanged. The ESG coefficients in columns (1) and (2) are significant at the 5% and 1% level, respectively. The coefficient on the interaction term between ESG and SOE in column (2) is negative and significant at the 10% level. In summary, the conclusions remain the same as before, and the results pass the robustness test.

**Table 6. Regression Results of the Robustness Test**

| | (1) | (2) |
|---|---|---|
| | FF | FF |
| ESG | 0.000** | 0.000*** |
| | (2.42) | (2.90) |
| ROE | 0.000*** | 0.000*** |
| | (3.21) | (3.22) |
| Size | 0.002 | 0.002 |
| | (0.94) | (1.09) |
| EBIN | 0.000*** | 0.000*** |
| | (2.97) | (3.06) |
| BM | 0.004*** | 0.004*** |
| | (5.16) | (5.20) |

| | | | |
|---|---|---|---|
| IND/YEAR | | control | control |
| ESG*SOE | | | -0.000* |
| | | | (-1.88) |
| _cons | | 0.089*** | 0.088*** |
| | | (4.69) | (4.61) |
| N | | 948 | 948 |
| $R^2$ | | 0.263 | 0.265 |

## 5. Further analysis

In financial markets, due to information asymmetry, firms may face financing constraints (Santos & Cincera, 2021), which refers to the difficulty of financing relative to investment opportunities. On the other hand, ESG can significantly mitigate information asymmetry by sending positive signals to the external world (Eliwa et al., 2021). Considering the above two points, we expect that there is a mediating role played by financing constrain in the process of ESG performance affecting financial flexibility, which we will analyze further in this section.

According to the research by Kaplan and Zingales (1997), this study utilizes the KZ index to measure Financing Constraints faced by firms. The KZ index consists of five financial ratios, and a higher KZ index indicates greater Financing Constraints for the firm. Then, to test the mediating role played by Financing Constraints between ESG performance and financial flexibility, this study adopts the mediation analysis approach proposed by Baron and Kenny (1986) and establishes Equations (3) and (4) based on Equation (1) as follows:

$$FC_{i,t} = \beta_0 + \beta_1 ESG_{i,t} + \beta_2 ROE_{i,t} + \beta_3 Size_{i,t} + \beta_4 EBIN_{i,t} \\ + \beta_5 BM_{i,t} + \sum IND + \sum YEAR + \varepsilon_{i,t} \quad (3)$$

$$FF_{i,t} = \beta_0 + \beta_1 ESG_{i,t} + \beta_2 FC_{i,t} + \beta_3 ROE_{i,t} + \beta_4 Size_{i,t} \\ + \beta_5 EBIN_{i,t} + \beta_6 BM_{i,t} + \sum IND + \sum YEAR + \varepsilon_{i,t} \quad (4)$$

In these two equations, $FC$ represents Financing Constraints as the mediating variable. Equation (3) examines whether ESG performance has an impact on Financing Constraints, while Equation (4) investigates the role of Financing Constraints as a mediating variable.

The results of the descriptive statistics of the financing constraints are shown in Table 2. The minimum value of financing constraints is -15.573 and the maximum value is 22.382, which represents that most of the firms face fair financing constraints.

In this section, this study first regresses ESG and financing constraints to test the effect of ESG on financing constraints. Then, the research regresses ESG and financial flexibility controlling for financing constraints. We verify whether financing constraints play a mediating role through the regression coefficients of the two variables. Finally, to ensure the validity of the findings, an endogeneity test and a robustness test are conducted following the same methodology as in the previous section.

**Table 7. Balance test for PSM of Equations (3) and (4)**

| Variable | Unmatched or Matched | (1) %bias | T test | (2) %bias | T test |
|---|---|---|---|---|---|
| ROE | U | -6.5 | -0.99 | -6.5 | -0.99 |
|       | M | 4.9   | 2.33  | 3.1   | 1.58  |
| Size  | U | 10.6  | 1.63  | 10.6  | 1.63  |
|       | M | -7.0  | -0.93 | 0.2   | 0.03  |
| EBIN  | U | -7.9  | -1.21 | -7.9  | -1.21 |
|       | M | -0.7  | -0.16 | -2.1  | -0.50 |
| BM    | U | 6.2   | 0.95  | 6.2   | 0.95  |
|       | M | -2.8  | -0.7  | -5.8  | -1.45 |
| IND/YEAR |   | control | control | control | control |
| FC    | U |       |       | -0.7  | -0.11 |
|       | M |       |       | -1.7  | -0.28 |

The balance test results of this PSM are shown in Table 7. The biases after matching are all less than 5% and it pass the balance test.

The regression results are shown in Table 8. Column (1) indicates that the coefficient of ESG on financing constraints is negative and significant at the 5% level, representing that ESG performance can decrease the financial constraints. Column (2) illustrates that the regression coefficient of ESG controlling for financing constraints is significant at the 10% level, while the regression coefficient of financing constraints is negative and significant at the 1% level, which means firms with lower financial constraints may have higher financial flexibility.

**Table 8. The Mediating Role of the Financing Constraints**

|        | (1)       | (2)      | (3)      | (4)       | (5)       | (6)      |
|--------|-----------|----------|----------|-----------|-----------|----------|
|        | FC        | FF       | FC       | FF        | FC        | FF       |
| ESG    | -0.009**  | 0.000*   | -0.020*  | 0.001*    | -0.005*** | 0.000    |
|        | (-2.27)   | (1.84)   | (-1.88)  | (1.95)    | (-2.76)   | (1.07)   |
| ROE    | -0.002*** | 0.000    | -0.001   | 0.000     | -0.002*** | 0.000    |
|        | (-3.58)   | (1.45)   | (-1.59)  | (0.01)    | (-3.55)   | (1.48)   |
| Size   | 0.290***  | 0.009*** | 0.191*** | 0.015***  | 0.290***  | 0.009*** |
|        | (8.05)    | (6.34)   | (2.89)   | (5.14)    | (8.07)    | (6.39)   |
| EBIN   | -0.000    | 0.000*** | 0.000    | 0.000***  | -0.000    | 0.000*** |
|        | (-0.67)   | (3.05)   | (1.05)   | (3.05)    | (-0.76)   | (3.04)   |
| BM     | 0.197***  | 0.008*** | 0.258*** | 0.010***  | 0.197***  | 0.008*** |
|        | (13.00)   | (13.65)  | (15.03)  | (10.96)   | (13.01)   | (13.75)  |
| IND/YEAR | control | control  | control  | control   | control   | control  |
| FC     |           | -0.025***|          | -0.030*** |           | -0.025***|
|        |           | (-20.31) |          | (-12.08)  |           | (-20.40) |
| _cons  | -2.384*** | 0.012    | -0.453   | -0.089**  | -2.633*** | 0.022    |
|        | (-5.17)   | (0.70)   | (-0.38)  | (-2.58)   | (-6.17)   | (1.38)   |
| N      | 941       | 941      | 229      | 217       | 947       | 947      |
| $R^2$  | 0.301     | 0.490    | 0.640    | 0.691     | 0.302     | 0.491    |

Columns (3) to (4) in Table 8 show the endogeneity test results, and Columns (5) to (6) show the robustness test results. From the coefficients and significance, it can be judged that the results passed both tests. In sum, the results suggest that ESG performance reduces financing constraints and thus increases firms' financial flexibility. It reflects the mediating role of financing constraints.

## 6. Conclusion

Despite the fact that COVID-19 has ended and the world economy has seen some recovery. However, the current world economy is still threatened by the deterioration of international relations, the continuation of wars, the climate crisis, and other factors. Under these circumstances, it is essential for companies to improve their financial flexibility, which allows them to have sufficient financial headroom to cope with economic risks and avoid problems such as bankruptcy. Companies with good ESG performance are able to send positive signals to the market and enhance the likelihood of obtaining financing. In addition, consumers are more likely to favor their products, leading to increased profits and enhanced financial flexibility. The above conditions motivate us to investigate the relationship between ESG performance and financial flexibility. The paper finds that good ESG performance significantly improves firms' financial flexibility. Moreover, the effect of ESG performance on financial flexibility will be weaker on SOEs compared to general firms. Further analysis finds that financing constraints play a partial mediating role in the process of ESG affecting financial flexibility.

Our findings have important implications for SOE governance, corporate operations, investment, and ESG disclosure: First, firms can improve their financial flexibility by disclosing and investing in ESG to cope with an increasingly uncertain environment. Second, governments and stock exchanges can further promote ESG regulations to facilitate the sustainable development of listed companies. Finally, investors can pay more attention to the ESG performance of firms and consider it as one of the investment decision factors to reduce risk.

Our study still has some limitations. First, the samples we chose are all from the Hong Kong stock market and lack data support from foreign firms. Second, for the ESG index, we chose the Refinitiv database, which has a higher level of internationalization, but did not choose the ESG evaluation system (e.g., Wind, Hang Seng, etc.) that is more in line with the Hong Kong and Chinese context. Third, since there is no standardized measure of financing constraints and financial flexibility, we used the KZ index and net operating cash flow to quantify these two variables and did not attempt other measures. Fourth, in addition to analyzing the overall sample, we analyzed the additional effect of SOEs, but did not delve into firms with other characteristics. Future research could be conducted using different variable measures and a more comprehensive sample, as well as an in-depth examination of firms with different characteristics to provide better insights.